\documentclass[prl,twocolumn,showpacs,letterpaper,showpacs,superscriptaddress]{revtex4}
\usepackage{graphicx,amsmath,amssymb,amsfonts,latexsym,color,dcolumn,bm,epsfig,subfigure}

\begin{document}
\title{Measurement of the short-range attractive force between Ge plates using a torsion balance}

\author{W. J. Kim}
\affiliation{Yale University, Department of Physics, P.O. Box 208120, New Haven, CT 06520, USA}

\author{A. O. Sushkov}
\affiliation{Yale University, Department of Physics, P.O. Box 208120, New Haven, CT 06520, USA}

\author{D. A. R. Dalvit}
\affiliation{Theoretical Division, MS B213, Los Alamos National Laboratory, Los Alamos, NM 87545, USA}

\author{S. K. Lamoreaux}
\affiliation{Yale University, Department of Physics, P.O. Box 208120, New Haven, CT 06520, USA}

\date{\today}

\begin{abstract}
We have measured the short-range attractive force between crystalline Ge plates, and found contributions from both the Casimir force and an electrical force possibly generated by surface patch potentials. Using a model of surface patch effects that generates an additional force due to a distance dependence of the apparent contact potential, the electrical force was parameterized using data at distances where the Casimir force is relatively small. Extrapolating this model, to provide a correction to the measured force at distances less than 5 $\mu$m, shows a residual force that is in agreement, within experimental uncertainty, with five models that have been used to calculate the Casimir force.

\end{abstract}


\maketitle

%%%%%%%%%%%%%%%

{\it Introduction.-}
The Casimir force has been a subject of great interest, both theoretically and experimentally, because it is a macroscopic manifestation of quantum vacuum effects \cite{Casimir,Milonni,Bordag}, and it can have significant effects in nanomechanical systems. Despite a number of successful measurements celebrated over the last decade \cite{Steve,Casexp}, early investigations of short-range forces \cite{Nancy,Stipe} report the possible systematic effects due to residual electrostatic forces. In particular, the observed variation in the effective contact potential, recently reported in \cite{KimPRARC} and later confirmed in \cite{Sven}, presents a problem of fundamental importance when setting limits to  predicted submicron corrections to Newtonian gravity in a Casimir force measurement \cite{Roberto}. The optical response of a particular sample under study must also be carefully considered \cite{Svetovoy}, as the accuracy of data on the optical properties of materials typically limits calculational accuracy to no better than 5\%. In principle, both electric and optical studies of a given sample are subject to a combination of various surface effects of electric origin, and it is important to understand these issues in order to accurately characterize fundamental interactions, such as the Casimir force and non-Newtonian gravity.

%%%%%%%%%%%%%%%%

\begin{figure}[t]
\includegraphics[width=0.8\columnwidth,clip]{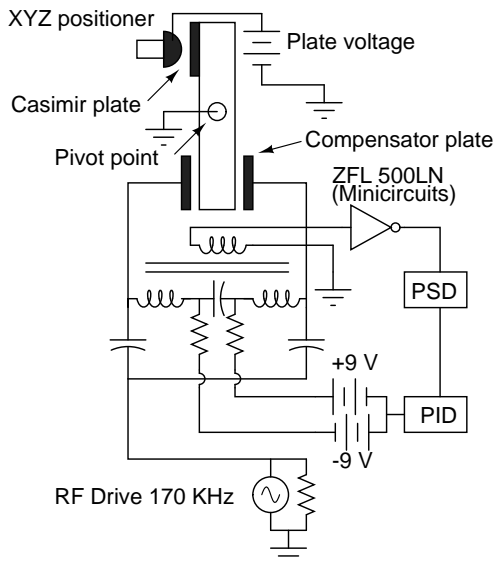}
\caption{Experimental setup of torsion balance (top-view). A pendulum body of length 15 cm hangs from a tungsten wire connected to a motorized rotation stage via the pivot point that is mounted on a support frame. The wire diameter is 25 $\mu$m, with length 2.5 cm, shorter than the previous experiment (66 cm) \cite{Steve} in order to minimize effects of tilt of the apparatus. At the bottom of the pendulum body (not shown in the figure) is a NdFeB magnet to damp the swinging modes of the pendulum at a natural frequency of 3 Hz. The mechanical assembly is covered by a glass bell jar (vacuum $5\times 10^{-7}$ torr) and is supported on a vibration isolation slab that has its foundation separate from the laboratory building.}
\label{fig1}
\end{figure}

{\it Our torsion balance set-up.-}
In this Letter, we present results of force measurements between crystalline Ge plates \cite{ispoptics} in a sphere-plane geometry.  Our apparatus, shown schematically in Fig. \ref{fig1}, is based on the design presented in \cite{Lamoreaux2} and improves on the apparatus described in \cite{Steve,fan}. On one side of a torsion pendulum a flat Ge plate is mounted, and approached by a Ge plate with a spherical surface, with radius of curvature $R=(15.10\pm0.05)$ cm, mounted on a Thorlabs T25 XYZ motion stage (8 nm resolution). When a force exists between these plates, the torsion body rotates and thereby generates an imbalance in capacitance on the other side of the pendulum, which carries a flat plate, situated in between two fixed ``compensator plates", that are attached to the support frame. An AC voltage is applied to the compensator plates, and the  capacitance imbalance creates an AC voltage that is amplified and sent to a phase sensitive detector (PSD), providing an error signal to a proportional-integral-differential  (PID) feedback circuit. A small correction voltage ($S_{\rm{PID}}$) is applied to the compensator plates keeping the system in equilibrium. The correction voltage is added to a large constant voltage $V_0 (\approx 9$V) to linearize the restoring force, $F\propto (S_{\rm{PID}}+V_0)^2\approx V_0^2+2V_0S_{\rm PID}$. This correction voltage provides a measure of the force between the Casimir plates and is recorded during the measurement.

The measured signal $S_{\rm{PID}}$ has contributions from several sources:
\begin{equation}
\label{PID}
S_{\rm{PID}}(d,V_{\rm a})=S_{\rm{DC}}(d\rightarrow\infty)+S_{r}(d)+S_{a}(d,V_{\rm{a}}),
\end{equation}
where $S_{\rm{DC}}$ is the force-free component of the signal at large distances, $S_{r}$ is the residual signal due to distance-dependent forces, such as Casimir-Lifshitz force, and $S_{\rm{p}}$ is the signal due to the electrostatic force in response to an applied external voltage $V_{\rm{a}}$. For the sphere-plane geometry, this latter signal
can be written in the proximity force approximation (PFA), valid when $d \ll R$, as $S_{a}(d,V_{\rm{a}})=\pi\epsilon_0 R(V_{\rm{a}}-V_{\rm{m}})^2/\beta d$, where $\beta$ is a calibration factor that converts $S_{\rm{PID}}$ in units of voltage to the actual units of force. The electrostatic signal is minimized ($S_{a}=0$)  when $V_{\rm{a}}=V_{\rm{m}}$, and the electrostatic minimizing potential $V_{\rm{m}}$ is then defined to be the contact potential between the plates.

A range of plate voltages $V_{\rm{a}}$ is applied, and at a given separation the response $S_{\rm{PID}}$ is fitted to a parabola
\begin{equation}
S_{\rm{PID}}(d,V_{\rm{a}})=S_0+k(V_{\rm{a}}-V_{\rm{m}})^2.
\label{para}
\end{equation}
The first two terms in Eq. (\ref{PID}) are absorbed in $S_0$ and represent the minimized signal when $V_{\rm{a}}=V_{\rm{m}}$. Repeating the parabola measurements shown in Fig. 2a, sequentially moving from the farthest to closest plate separations, enables us to inspect the $d$ dependence of the fitting parameters $k(d)$, $V_{\rm{m}}(d)$, and $S_0(d)$. The procedure outlined here was first implemented as a calibration routine in \cite{Iannuzzi} and more recently in \cite{KimPRARC} in an effort to detect a distance dependence of $V_{\rm{m}}$.

\begin{figure}[t]
\includegraphics[width=1.0\columnwidth,clip]{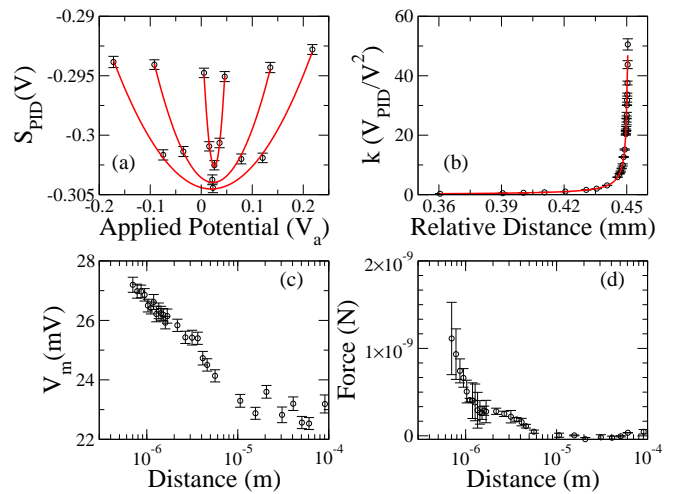}
%\vspace{0.5cm}
\caption{Description of procedure for force analysis. (a) A single parabola measurement at a given distance is acquired by sweeping a range of voltage differentials applied to the plates. The procedure is repeated at decremental distances from 150 $\mu$m down to 500 nm, completing a single experimental run. The parabolic curves shown here represent only three distances, 100 $\mu$m, 20 $\mu$m, and 1 $\mu$m for clarity. (b) Curvature coefficients of the parabola $k$ versus relative distances were fitted to the $1/d$ electric force to provide the voltage-to-force conversion factor $\beta$ as well as the absolute distance obtained from an asymptotic limit. (c) The force-minimizing potential $V_{\rm{m}}$ changes with distance, varying approximately as $\log d$. (d) The residual force at the minimizing potential is plotted against distance, after subtracting the DC offset $S_{\rm{DC}}$ and multiplying it by  $\beta$. The maximum force gradient for feedback system stability is 5 nN/$\mu$m, limiting the minimum distance to 500 nm.}
\label{fig2}
\end{figure}

As the gap between the plates is reduced, the parabola curvature $k$ rapidly increases as shown in Fig. \ref{fig2}b. These curvature values are fitted to $k(d)=\alpha/d$, the expected dependence for the plane-sphere geometry, where the absolute distance $d\equiv d_0-d_{\rm{r}}$ is defined in terms of the asymptotic limit $d_0$ and the relative distance $d_r$ recorded during a parabola measurement. The conversion factor $\beta$ is then obtained through $\alpha\equiv\pi\epsilon_0R/\beta$. Obviously, $\alpha$ can be also used to determine the absolute distance through  $d=\alpha/k$, implying a significant correlation of $\alpha$ with $d_0$. Consistency between these two methods of distance determination reflects validity of the use of the $1/d$ power law as implied by a value of $\chi^2_0$ close to unity for our data set. Fig. \ref{fig2}c shows the electric potentials $V_{\rm{m}}$ at minima of the parabola curvatures plotted versus $d$, indicating the distance-dependent minimizing potential $V_{\rm{m}}(d)$, a behavior that has been observed in other experiments \cite{KimPRARC,Sven}.

To see the trend in $V_{\rm m}(d)$ more clearly and to determine short-range forces with higher statistical accuracy, we have repeated 200 times the experimental sequence described in Fig. \ref{fig2}, yielding a total of 5800 data points. Each group of five data points taken at a given fixed distance with varying applied potential are used to determine the three parabola parameters discussed above, in addition to the force and distance. The mean value of the calibration factor after analyzing all data is $\beta=(1.35\pm0.04)\times 10^{-7}$~N/V. Both the asymptotic limit $d_0$, shown in Fig \ref{fig2}b, and the DC offset of the PID signals $S_{\rm{DC}}$ drift slightly during a run. The uncertainty in position is roughly 10 \% at a given distance and about 50 nm at the typical closest gap separation, consistent with the actuator minimum displacement of 40 nm. The DC offset drift has been corrected by monitoring $S_{\rm{PID}}$ before and after each consecutive run and applying a linear correction.

%%%%%%%%%%%%

\begin{figure}[t]
\includegraphics[width=1.0\columnwidth,clip]{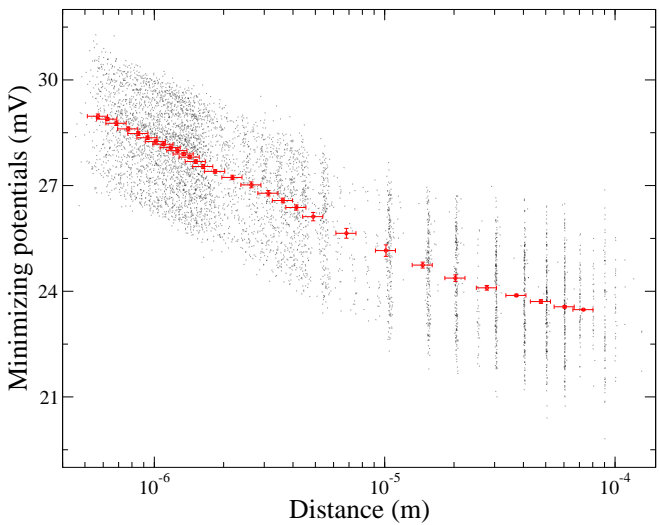}
\caption{The value of the force minimizing potential as a function of plate separation. The red points are the average of the
data.  Our measurement reveals a slow rise of the minimizing potential as the plates approach each other, of order of 6 mV over 100 $\mu$m. This variation in our data set shows a similar trend observed in a recent measurement  \cite{Sven} where the variation of 6 mV over 1 $\mu$m is reported between Au coated samples, significantly larger than the value found in our study.}
\label{fig3}
\end{figure}

{\it Varying minimizing potential.-}
An outstanding feature of our data is the distance variation of the applied voltage $V_{\rm{m}}$ that minimizes the force, as clearly shown in Fig. \ref{fig3}. It must be recognized that this variation can lead to an extra force of electrical origin, as demonstrated in  \cite{Lamoreauxcont}. However, the model used in \cite{Lamoreauxcont} assumes that the variation in the minimizing potential is due to a varying contact potential, specifically modeled as a voltage source in series with the plates. The varying minimizing potential observed in our data is more likely due to large-scale gradients in the contact potential across the surface of the plates, due to, for example, polishing stresses or the curvature of the spherical surface plate changing the crystal plane orientation at the surface. Such variations have been observed for many materials  \cite{Robertson,INFN},  with typical large scale fluctuations on the order of a few mV. The variation in the apparent contact potential is due to the effective averaging area changing as the curved and flat surfaces of the plates are brought together. A numerical analysis \cite{Note} of a wide range of surface potential variations shows that the variation of $V_{\rm{m}}(d)$ leads to an electrostatic force of the form  $F^{\rm el}_{r_1}(d)= \pi R\epsilon_0[V_{\rm{m}}(d)+V_1]^2 / d$, where $V_1$ is a constant offset parameter at large distances that can be determined from the experimental data. The origin of this effect is due to the plate curvature, together with large scale variations in the surface contact potential. Note that although the parabola measurement {\it minimizes} the electrostatic force across the plates, it does not necessarily {\it nullify} all the electric forces that possibly exist.

%%%%%%%%%%%%%%%%
{\it Random small-scale patches.-}
In addition to large scale gradients in the surface potential, there can be small-scale (i.e., much smaller than the plate diameter) random fluctuations in the surface potential associated with strains, irregularities, and impurities. It is straightforward to show that the electrostatic energy per unit area between two flat plates with random patch voltages is \cite{Speake,Note}
\begin{equation}
E_{\rm patch}(d)=\frac{\epsilon_0 \pi V_{\rm rms}^2}{4 d}  \int_0^\infty  du S(u/d) \frac{e^{-2u}}{\sinh^2(u)},
\label{patch_residual}
\end{equation}
where $V_{\rm rms}$ is the rms value of the random patch voltages. For simplicity, we have assumed isotropic patches with surface correlation functions $\langle V_{k,i} V_{k',j} \rangle = V^2_{\rm rms} S(k)  \delta_{i,j} \delta(k-k')$, where $i,j=1,2$ denote the plates, and $S(k)$ is the unity-normalized spectral density. The residual electrostatic force between the sphere and the plane due to these patches can be obtained from PFA as $F^{\rm el}_{r_2}(d)=2 \pi R E_{\rm patch}(d)$.  For example, for random-voltage patches of radius $\lambda$ uniformly distributed on the surfaces, the spectral density is $S(k) \approx \sin k \lambda / \pi^2 k^2$. It is easy to see that in the limit $\lambda \gg d$ the residual patch force in the sphere-plane geometry scales as $F^{\rm el}_{r_2}(d)=\pi R\epsilon_0 V^2_{\rm{rms}}/d$ \cite{Note}.

%%%%%%%%%%%%%%%%%

{\it Electrostatic residual force.-}
We fit the data of the residual force at the minimizing potential (Fig 2.d) with a force of electric origin $F^{\rm el}_r = F_0+F^{\rm el}_{r_1} + F^{\rm el}_{r_2} =F_0+\pi R \epsilon_0 \{[V_{\rm{m}}(d)+V_1]^2 + V^2_{\rm rms} \}/d$, where $F_0$ is an offset parameter at large distances. A least-squares fit of the observed force using data for $d>5$ $\mu$m (a regime where the Casimir force should be vanishingly small) and the measured $V_{\rm{m}}(d)$, while leaving $F_0$, $V_1$, and $V_{\rm{rms}}$ as adjustable fit parameters, yields an excellent description of the observed large distance force \cite{powerlaw}, as shown plotted with the data in Fig. 4. Including data at shorter distances ($d<5$ $\mu$m) causes a significant fit deviation, indicating an interference with the actual Casimir force which is highly nonlinear at short distances. A similar long-range force has been previously observed in the measurement of van der Waals interaction and the corresponding correction is applied to the data based on work function anisotropies and their related patch charges \cite{Nancy}.

%%%%%%%%%%%%%%%%%

{\it Casimir residual force.-}
We have tested our data for  the presence of a residual Casimir force $F_r^{\rm Cas}(d)$ between the Ge plates, which we have computed in PFA from the plane-plane Casimir-Lifshitz energy, $F^{\rm Cas}_r(d) = 2 \pi R E^{\rm Cas}_{pp}(d)$. We have calculated the corresponding reflection coefficients using five different theoretical models for the Ge plates \cite{Diego, Galina}: ideal dielectric;  ideal dielectric + Drude conductivity corrections; ideal dielectric + plasma conductivity corrections; quasi-static Debye-H\"uckel screening model; and charge drift model. Fig. 4 (bottom) shows the experimental data for the residual force after subtraction of the joint (contact potential and patch potential) electrostatic forces, and the theory curves for the Casimir-Lifshitz force between Ge plates at $T=300$ K for these five theoretical models. The error bars take into account all statistical uncertainties (2-3 \%) as well as fitting uncertainties from the electrostatic force analysis (10\%). Within experimental uncertainty, our data agrees well with the theoretical predictions for the Casimir force, but it is not of sufficient accuracy to distinguish among the different models for the Ge plates. High precision measurements of the Casimir force require a careful evaluation of the electrostatic effects considered in detail for the first time in our present study.
\begin{figure}[t]
\includegraphics[width=1.0\columnwidth,clip]{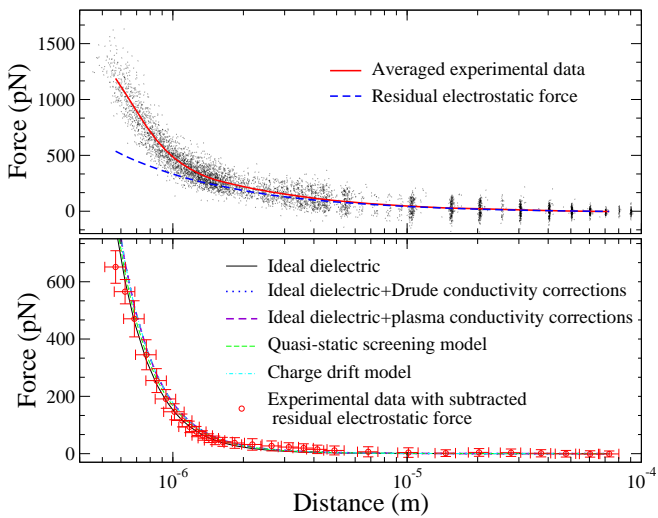}
\caption{(top) Experimental data from parabola measurements along with a solid line representing an average. The joint (contact potential and patch potential) electrostatic residuals is modeled by  $F^{\rm el}_r=F_0+ \pi R\epsilon_0 \{ [V_{\rm{m}}(d)+V_1]^2 + V^2_{\rm rms} \}/d$, which is fit to our data points, yielding $F_0=(-11\pm2)\times10^{-12}$ N, $V_1=(-34\pm 3)$ mV, $V_{\rm{rms}}=(6\pm2)$ mV, and $\chi^2_0=1.5$. (bottom) Experimental data for the remaining force after subtraction of electrostatic correction $F_r^{\rm{el}}$, together with theoretical Casimir forces computed using five models described in \cite{Diego,Galina}. $\chi^2$ for $d < 5$ $\mu$m is approximately unity with the corrected data for all of the models.}
\label{fig4}
\end{figure}

{\it Conclusions.-} We have performed measurements of the short-range force between Ge plates in the sphere-plane geometry, and have observed that the potential $V_{\rm{m}}$ that minimizes the electrostatic force depends on the gap between the plates. We have considered two contributions of electric origin present in the residual data for the force. The first contribution is due to large-scale variations in the contact potential along the surface of the plates, that leads to the gap-dependent minimizing potential and, as a result, to an electrostatic force proportional to $(V_{\rm{m}}(d)+V_1)^2/d$.  The second contribution can be modeled as arising from potential patches on the surfaces that, in the case when they have typical sizes much smaller that the plate diameters and much larger than the plate separation, leads to a further electrostatic force proportional to $V_{\rm rms}^2/d$. We have fitted our experimental data at large distances ($d>5$ $\mu$m, where the Casimir force is expected to be negligible) with these two electrostatic force effects, and found we could establish good agreement between our model and the experimental data. Furthermore, we have subtracted these forces from the data at short separations ($d<5$ $\mu$m) and found a residual force that is in agreement with the theoretical predictions for the Casimir-Lifshitz force between Ge plates. Our measurements do not have enough accuracy to distinguish between the different theoretical models used to characterize the optical properties of the Ge plates. Future measurements are deemed necessary in light of our discussion, in particular to better understand the physical origins of the observed electrostatic forces. We are currently exploring similar surface effects in a pair of Au samples. We acknowledge support from Yale University for the construction of the experimental apparatus and data acquisition, and from Los Alamos LDRD program. We thank G. Klimchitskaya for useful discussions.

%%%%%%%%%%%%%%%%%%%%%


\begin{thebibliography}{99}
\bibitem{ispoptics}
We chose Ge for our measurements because it is among the purest materials available, optical components made  of intrinsic crystalline Ge are readily available, and its properties are well-tabulated. (The lenses used are from ISP optics, GE-PX-25-50.)
\bibitem{powerlaw}
Incidentally, the long range residual force can also be fitted with a power law $F^{\rm el}_r=a_0+a_1/d^n$, yielding $a_1=(-1.82\pm0.6)\times10^{-11}$ N, $a_1=(1.6\pm0.2)\times10^{-14}$ N $\rm{m}^{0.72}$, and $n=0.72\pm0.2$, with $n$ consistent with a calculation, presented in \cite{Speake}, of the effects of surface adsorbates fort the sphere-plane geometry.  However, these effects do not directly lead to a varying $V_m$.
\end{thebibliography}
\end{document}